\documentclass[aps,prl,reprint,superscriptaddress,longbibliography]{revtex4-1}
\usepackage{graphicx}
\usepackage{amssymb}
\usepackage{amsmath}
\usepackage{appendix}
\usepackage{comment}
\usepackage{hyperref}
\usepackage{color}
\usepackage{soul}

\usepackage{tikz}       % Feynman diagrams
\usepackage{tikz-feynman}
\makeatletter
\tikzfeynmanset{compat=\tikzfeynman@version@major.\tikzfeynman@version@minor.\tikzfeynman@version@patch}
\makeatother

\newcommand{\ak}[1]{{\color{magenta} #1}}

\begin{document}

\title{Spin current cross-correlations as a probe of magnon coherence}

\author{Scott A. Bender}
\affiliation{Utrecht University, Princetonplein 5, 3584 CC Utrecht, The Netherlands}

\author{Akashdeep Kamra}
\affiliation{Center for Quantum Spintronics, Department of Physics, Norwegian University of Science and Technology, Trondheim, Norway}
\affiliation{Department of Physics, University of Konstanz, D-78457 Konstanz, Germany}

\author{Wolfgang Belzig}
\affiliation{Department of Physics, University of Konstanz, D-78457 Konstanz, Germany}

\author{Rembert A. Duine}
\affiliation{Utrecht University, Princetonplein 5, 3584 CC Utrecht, The Netherlands}
\affiliation{Department of Applied Physics, Eindhoven University of Technology,
P.O. Box 513, 5600 MB Eindhoven, The Netherlands}
\affiliation{Center for Quantum Spintronics, Department of Physics, Norwegian University of Science and Technology, Trondheim, Norway}

\begin{abstract}

 Motivated by the important role of the normalized second order coherence function, often called $g^{(2)}$, in the field of quantum optics, we propose a method to determine magnon coherence in solid-state devices. Namely, we show that the cross-correlations of pure spin-currents injected by a ferromagnet into two metal leads, normalized by their dc value, replicate the behavior of $g^{(2)}$ when magnons are driven far from equilibrium. We consider two scenarios: driving by ferromagnetic resonance, which leads to the coherent occupation of a single mode, and driving by heating of the magnons, which leads to an excess of incoherent magnons. We find an enhanced normalized cross-correlation in the latter case, thereby demonstrating bunching of nonequilibrium thermal magnons due to their bosonic statistics. Our results contribute to the burgeoning field of quantum magnonics, which seeks to explore and exploit the quantum nature of magnons.

\end{abstract}

\maketitle

%\akcom{I have attempted to write an introduction which should give a decent overview and motivation to the spintronics community which may not be well-versed with quantum optics.}
%\akcom{We might want to reduce the number of equations in the paper. From what I remember, PRL guidelines suggest limiting the number of equations to 12 or so.}

%---------------------------------------------Introduction------------------------------------------- %

{\it Introduction.} --- In the early years of quantum mechanics, the drive towards demonstrating its consistence with classical physics led Schroedinger to examine a class of wavefunctions that replicates the classical dynamics of a harmonic oscillator~\cite{Nieto1997}. In the language of second quantization, these turned out to represent the eigenstates of the annihilation operator~\cite{Glauber1963,Sudarshan1963}. These have come to be known as `coherent states' and are often termed the `most classical' in nature. Remarkably, the purely quantum phenomenon of Bose-Einstein (BE) condensation results in a condensate characterized by a coherent state wavefunction. Thus, phenomena like superconductivity and superfluidity can largely be described in terms of and bear an intricate relation to coherent states.

Characterizing and quantifying the ``quantumness'' of photonic states is one of the central themes in the field of quantum optics~\cite{Walls2008,Gerry2004}. A measure particularly relevant for condensation phenomena is the so-called normalized second-order temporal coherence function, typically denoted by $g^{(2)} (\tau)$:
\begin{align}\label{eq:g2}
g^{(2)} (\tau) = & \frac{\left \langle : \hat{I}(t) \hat{I}(t + \tau) : \right \rangle}{\left \langle \hat{I}(t)  \right \rangle \left \langle \hat{I}(t + \tau) \right \rangle},
\end{align}
where $\hat{I}$ is the optical intensity magnitude operator, $:~:$ represents normal ordering of the photon ladder operators, and $\langle ~ \rangle$ denotes the expectation value. For a coherent state (e.g. a laser), the numerator in Eq. (\ref{eq:g2}) factorizes, and $g^{(2)} (\tau) = 1$. $g^{(2)} (\tau)$ also quantifies the relative probability of detecting two photons separated by a time lag $\tau$. Specifically, $g^{(2)} (0)$ gives the rate of simultaneous detection of two photons in a given optical state normalized by the analogous rate exhibited by a coherent or classical light field. A single-mode thermal light field exhibits the so-called photon bunching effect, corresponding to $g^{(2)} (0) = 2$, which is indicative of the bosonic nature of the photons. The concept of bosonic bunching has been extended beyond quantum optics in more recent years; an analogue of $g^{(2)}(\tau)$ has been measured in a system of ultracold trapped atoms, demonstrating temporal atomic bunching and BE condensate coherence \cite{Guarrera:2011by,*Naraschewski:1999ic}.  

While quantum optics is, by now, a mature field with several experiments on optical BE condensates~\cite{Greveling:2018ke,*Klaers2010}, non-classical states of light~\cite{Slusher1985}, photon coincidence statistics~\cite{Short1983,*Zou1990} and so on, similar advances have just begun to be made for bosonic excitations in magnets - magnons~\cite{Bauer2012,*Kruglyak2010,*Chumak2015,*Sonin2010,Kamra2016,*Kamra2017}. Recent advances~\cite{Hirsch1999,*Saitoh2006,*Valenzuela2007,*Kajiwara2010} in manipulating and detecting these excitations have enabled exciting fundamental physics in magnetic systems with the potential for technological applications~\cite{Zutic2004}. Observation of a non-equilibrium magnon BE condensate has been claimed on the basis of high occupancy recorded in the lowest energy magnon mode, when a large number of higher energy magnons are pumped into the system~\cite{Demokritov2006,Chumak2015}. In these systems, magnon coherence has been demonstrated by various optical techniques \cite{Demidov:2008eu,*Jackl:2017hz}. The same system seems to exhibit a superfluid-like flow of spin current providing further evidence towards these claims~\cite{Bozhko2016}, which, however, are not uncontested~\cite{Sonin2016,*Bozhko2016B}. Existence of robust spin-superfluidity with a different physical origin has been postulated in axially-symmetric ferromagnets as well as antiferromagnets \cite{Sonin:2010do,Skarsvag:2015cr,Takei:2014dc,Flebus:2016kv}. However, a direct signature of magnon coherence measurable in all-solid-state systems, where optical probes may not be feasible, is still lacking.

In this Letter, we give a proposal for the solid-state detection of magnon state coherence, adapting concepts from quantum optics to metal/magnet hybrid structures. We show that metallic contacts, such as those used in nonlocal magnon transport experiments \cite{Cornelissen:2015cz}, can play the role of coherence-sensitive ``detectors" of magnon emission, in analogy with the photon detectors in quantum optical experiments. A key difference between quantum optical and spintronic systems lies in the fact that light, due to its long wavelength, weak interaction with surroundings, and ability to propagate in vacuum, exhibits coherence over large distances. This eliminates the need for the explicit inclusion of detectors in its description. In solid-state systems, we find it prudent to develop the coherence theory of magnons for a specific realization of detection integrated into the device structure. Nevertheless, the ideas developed herein are general and can be adapted to a different detection scheme. Furthermore, our findings are suggestive of employing heat currents as a probe into phonon coherences in an analogous manner.

\begin{figure}[pt]
\includegraphics[width=0.6\linewidth,clip=]{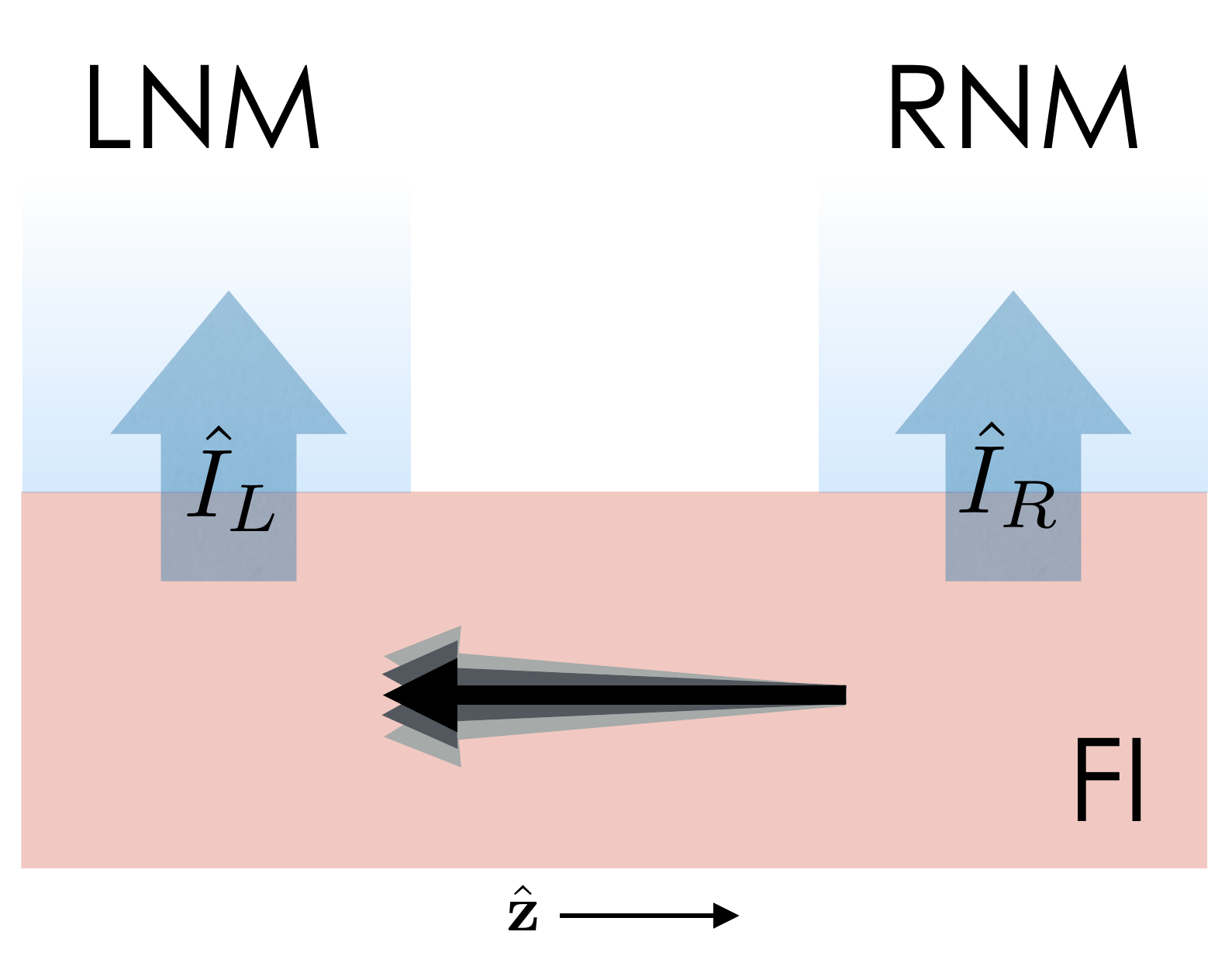}
\caption{Magnetic heterostucture, with ferromagnetic insulator (FI) and metallic leads (LNM and RNM). The arrow in FI depicts its total spin. Spin currents $\hat{I}_L$ and $\hat{I}_R$ are injected into the LNM and RNM, respectively, when the FI is excited.}
\label{sch}
\end{figure}

{\it Normalized spin current cross-correlation.} --- We consider a ferromagnetic insulator (FI) in contact with two non-magnetic metal leads (LNM and RNM) (e.g. see Fig. \ref{sch}), into which spin current is injected and can be measured via the inverse spin Hall effect~\cite{Hirsch1999,Saitoh2006}. We define the normalized spin current cross-correlation $c^{(2)}(\tau)$:
\begin{align}\label{eq:c2}
c^{(2)} (\tau) \equiv & \frac{\frac{1}{2}\left \langle  \{ \hat{I}_L(t), \hat{I}_R(t + \tau) \} \right \rangle}{\left \langle \hat{I}_L(t)  \right \rangle \left \langle \hat{I}_R(t + \tau) \right \rangle},
\end{align}
where $\hat{I}_{l}$ with $l=L,R$ are operators corresponding to the spin currents injected into LNM and RNM, respectively, and $\{\dots\}$ denotes an anticommutator. A key feature of this measure, distinguishing it from $g^{(2)} (\tau)$, is that the spin currents may be positive or negative. In this sense, LNM and RNM at finite temperatures may be considered as ``non-ideal detectors'' which may also emit magnons back into the FI, in addition to absorbing (detecting) them from it. The optical detectors, in contrast, are designed to predominantly absorb, and hence detect, photons. Thus, the positivity of intensity in Eq. (\ref{eq:g2}) is maintained in this case. 

\begin{figure}[pt]
\includegraphics[width=\linewidth,clip=]{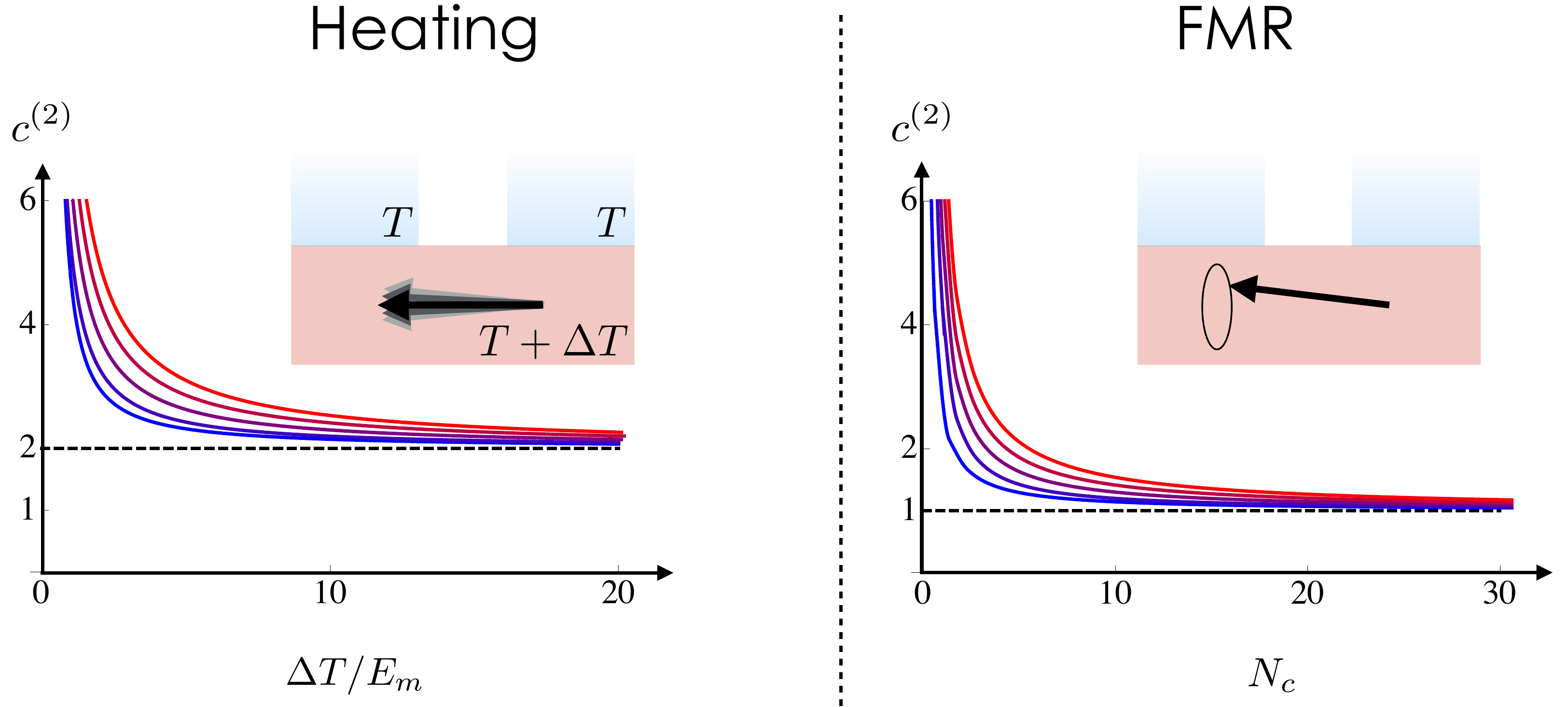}
\caption{Normalized spin-current cross-correlation $c^{(2)}$ for FI driven via heating (left) and FMR (right). The coherent mode number $N_c=(\chi h)^2$ is assumed to be driven by a circularly polarized FMR field $h$, with $\chi$ as the susceptibility. Solid lines, blue to red, represent electronic temperatures $T/E_m=0.01,2.5,5,7.5,10$.}
\label{results}
\end{figure}

In the present work, we evaluate $c^{(2)} \equiv c^{(2)} (0)$ considering a single magnon mode driven into (i) a coherent state and (ii) a thermal state with temperature higher than the metal leads. We find that at sufficiently large drives, $c^{(2)}$ emulates the behavior expected from $g^{(2)} \equiv g^{(2)} (0)$ (Fig. \ref{results}), i.e. it approaches 1 and 2 for cases (i) and (ii) respectively, demonstrating it to be a valid coherence measure. The need for strong driving of the FI arises because magnon detection via the metal leads is rendered non-ideal by the processes in which the metal leads emit magnons into the FI. At sufficiently large drives, quantified by the temperature of the metal leads, magnon absorption by the metal lead detectors dominates over emission, and the detection process is efficient and similar to its optical counterpart.

\begin{comment}

\section{Introduction}
Suggestions for outline of introduction (we may need to change some equations in the body of the paper depending on what we decide to define here in the introduction):

-define/discuss $g^{(2)}=\langle \hat{\phi}^\dagger \hat{\phi}^\dagger \hat{\phi}  \hat{\phi}\rangle/|\hat{\phi}^\dagger \hat{\phi}|^2$ / role in other systems

-define spin current noise 
\begin{align}
\mathcal{S}(t-t')= \frac{1}{2}\lim_{t'\rightarrow t}\langle\{\delta \hat{I}_L(t),\delta\hat{I}_R(t')\} \rangle\\
=\int \frac{d\omega}{2\pi }\mathcal{S}(\omega)e^{-i\omega(t-t')}
\end{align}
with $\mathcal{S}(\omega)$ as typically experimentally measured quantity. Define correlator $\mathcal{C}\equiv \frac{1}{2}\lim_{t'\rightarrow t}\langle\hat{I}_L(t)\hat{I}_R(t') +\hat{I}_R(t')\hat{I}_L(t)\rangle=I_L I_R+\mathcal{S}$; state how $\mathcal{C}=I_L I_R$ for coherent modes.
 
-Define normalized correlator as $c=\mathcal{C}/I_L I_R=1+S/I_L I_R=1+s$; state how $c=1$ for coherent and $c=2$ for incoherent

-Discuss heating of FI versus FMR results- see Fig 2

\end{comment}

{\it Electron-magnet coupling.} ---  We consider the FI to be axially-symmetric around the $z$-direction. Accordingly, we take the equilibrium macrospin to be oriented in the $-\mathbf{z}$ direction, so that excitations thereof carry spin in the $+\mathbf{z}$ direction. For simplicity, in order to contrast coherent and incoherent spin excitations of the FI, we allow for small angle dynamics of the macrospin $\mathbf{S}$ only, neglecting micromagnetic degrees of freedom.  Such excitations are conveniently parametrized by the Holstein-Primakoff transformation:
\begin{equation}
\hat{S}_{z}\equiv \hat{\phi}^{\dagger}\hat{\phi}-S,\, \,\, \hat{S}_{-}\equiv \hat{S}_{x}-i\hat{S}_{y}=\sqrt{2S-\hat{\phi}^{\dagger}\hat{\phi}}\hat{\phi}\, ,
\label{hp}
\end{equation}
where $S$ is the integer macrospin of the FI, and $\hat{\phi}$ and $\hat{\phi}^\dagger$ are respectively magnon annihilation and creation operators, subject to the bosonic commutation relations $[\hat{\phi},\hat{\phi}^\dagger]=1$. The operator $\hat{\phi}$ has two components:
\begin{equation}
\hat{\phi}=\hat{\varphi}+\varphi\,.
\end{equation}
The first term, $\hat{\varphi}$, corresponds to incoherent fluctuations, with $\langle \hat{\varphi} \rangle \equiv 0$, while the second, $\varphi$, is a c-number corresponding to a coherent magnon. Only  $\varphi$ contributes to the ensemble-averaged transverse dynamics. Assuming small amplitudes for $\varphi$ and $\hat{\varphi}$, we obtain $\langle S_x \rangle=\sqrt{2S}\mathrm{Re}[\varphi]$ and $\langle S_y \rangle=-\sqrt{2S}\mathrm{Im}[\varphi]$. 

The evolution of the heterostructure spin dynamics is governed by the Hamiltonian $\mathcal{H}=\mathcal{H}_m+\mathcal{H}_e+\mathcal{H}_J$. The uncoupled FI magnon Hamiltonian is given by $\mathcal{H}_m=E_m \hat{\phi}^\dagger \hat{\phi} $, with $E_m$ as the magnon gap or, equivalently, $\hbar$ times the ferromagnetic resonance frequency. The uncoupled normal metal electron Hamiltonian is: $\mathcal{H}_e=\sum_{l=L,R}\mathcal{H}_l$, where $\mathcal{H}_l=\sum_{k \sigma} \epsilon_k \hat{b}_{l k\sigma }^\dagger \hat{b}_{ lk\sigma }$, with $\hat{b}_{lk\sigma}$ as a $\sigma$-spin annihilation operator for an electron in the $l=L,R$ lead with quantum number $k$~\footnote{For simplicity, we assume identical leads, so that electrons in both leads have identical dispersions.}. Magnons in the FI and the spins of electrons in the normal metals are coupled by exchange at the metal-magnet interfaces, which is captured by hopping of spin between magnons and electron-hole excitations in the normal metals: $\mathcal{H}_J=\sum_{K} J_K \hat{\phi}^\dagger ( \hat{L}_{K}+ \hat{R}_{K})+H.c.$ \cite{Bender:2015jx}. Here $J_K$ is the effective exchange interaction for both interfaces, while $ \hat{L}_{K}^\dagger =\hat{b}^\dagger_{Lk\uparrow} \hat{ b}_{L\tilde{k}\downarrow}$ and $ \hat{R}_{K}^\dagger =\hat{b}^\dagger_{Rk\uparrow} \hat{ b}_{R\tilde{k}\downarrow}$ are creation operators for up-electron/down-hole pairs in the left and right leads, respectively, with $K=k,\tilde{k}$ as a collective index. 

The coupling $\mathcal{H}_J$ gives rise to spin currents through the left and right interfaces. The operator expression for each of these currents is obtained from the Heisenberg equation of motion for the electron spin density $\hat{s}_l=\sum_K  (\hat{b}^\dagger_{lk\uparrow} \hat{b}_{lk\uparrow}-\hat{b}^\dagger_{lk\downarrow} \hat{b}_{lk\downarrow})$ in the respective lead. The current in the left lead is:
\begin{equation}
\nonumber
\hat{I}_L= \frac{i}{2} [\hat{\mathcal{H}}_J,\hat{s}_l]=\sum_{K} (\hat{L}^\dagger_K-\hat{L}_K )\, .
%\label{IL}
\end{equation}
Similarly, the current in the right lead is: $\hat{I}_R=\sum_{K} (\hat{R}^\dagger_K-\hat{R}_K )$.

%For simplicity, we assume strong spin relaxation in the metallic leads and no electronic driving, so that any spin order in the LNM and RNM vanishes.

In equilibrium, the average currents $I_L=\langle \hat{I}_L \rangle$ and $I_R=\langle \hat{I}_R \rangle$ vanish. The exchange Hamiltonian also gives rise to correlations between LNM and RNM electrons. To zero$^{th}$ order in $\hat{\mathcal{H}}_J$, such correlations are destroyed by coupling to the environment, and the equilibrium density averages are unaffected by the exchange coupling. Thus, for LNM and RNM electrons in equilibrium, $\langle \hat{b}_{lk\sigma}^\dagger \hat{b}_{l'k'\sigma'}\rangle=\delta_{ll'}\delta_{kk'}\delta_{\sigma\sigma'} n_{F }(\epsilon_{k})\,$ where $n_{F}(\epsilon)=1/(1+e^{(\epsilon-\epsilon_F)/T})$ is the Fermi-Dirac distribution with common electronic temperature $T$ and Fermi energy $\epsilon_F\gg T$.  As phase coherent dynamics vanish in equilibrium, the $z$-component of the incoherent spin density under equilibrium conditions is: $\langle \hat{\varphi}^\dagger \hat{\varphi}\rangle=N=N_i$, with $N_i=1/(e^{E_m /T_m}-1)$ as the Bose-Einstein distribution for magnon energy $E_m$ and temperature $T_m$.

{\it Driving the magnet.} --- We assume the normal metals to be ideal heat and spin sinks, so that the electrons there are described by equilibrium conditions discussed above. We consider two methods of directly driving the FI. First, ferromagnetic resonance (FMR) can coherently excite magnetic dynamics. Under the influence of a microwave field with frequency $\omega$, a coherent excitation described by $\varphi(t)=\varphi e^{i\left(\omega t+\theta \right)}$ is created. Here the $U(1)$ symmetry of the FI is explicitly broken, as the precessional phase $\omega t+\theta$ is determined by the applied microwave. As a second driving scheme, incoherent magnetic dynamics can be excited beyond equilibrium by heating the FI to a temperature $T_m$ higher than that of the metal leads, for example by heating with a laser pulse.  Thus, in the presence of one or both types of drives, the nonequilibrium magnon density, to zero$^{\mathrm{th}}$ order in $\hat{H}_{J}$ becomes $\langle \hat{\varphi}^\dagger \hat{\varphi}\rangle=N=N_c+N_i$, where $N_c=\varphi^* \varphi$. The steady-state current resulting from either type of drive has been calculated to second order in the exchange coefficient $J_K$ \cite{Bender:2015jx}:
\begin{equation}
\langle \hat{I}_L \rangle=\langle \hat{I}_R \rangle=I_L=I_R=2\pi D^2   J^2 E_m \left(N-N_{\mathrm{NM}} \right)\, ,
\label{totcurrent}
\end{equation}
where $N_{\mathrm{NM}}=1/(e^{E_m/T}-1) $ is the Bose-Einstein distribution function describing electron-hole excitations in the leads.  The Fermi-surface averaged square of the exchange interaction is given by $J^2\equiv \sum_{K} \left| J_{K}\right|^2 \delta(\epsilon_k-\epsilon_F) \delta(\epsilon_{\tilde{k}}-\epsilon_F)/D^2$, with $D$ as the electronic density of states at the Fermi energy $\epsilon_F$ in the metal leads. Thus we see that it is not possible to distinguish the coherent and incoherent magnons from the spin currents $I_L$ and $I_R$ alone, since these depend on the total number of magnons $N=N_c+N_i$. Instead, we turn to the cross-correlations of the spin currents.

{\it Spin current cross-correlations.} ---  We now investigate the current-current cross correlator $C(\tau)=\frac{1}{2}\langle \{ \hat{I}_L(t), \hat{I}_R(t + \tau) \} \rangle$ to obtain $c^{(2)}(\tau)$ in Eq.~(\ref{eq:c2}).  One can see directly how $C(\tau)$ encodes information about magnon fluctuations. When the magnons are completely coherent ($\hat{\phi}=\varphi$) and magnon fluctuations can be neglected, the coupling Hamiltonian $\mathcal{H}_J$ does not give rise to correlations between left and right electrons. Here, $C(\tau)=\langle \hat{I}_L(t)\rangle \langle \hat{I}_R(t+\tau) \rangle$ factorizes, yielding $c^{(2)}(\tau)=1$. If, however, incoherent fluctuations of the magnon operators are taken into account, then cross-correlations give rise to
\begin{equation}
c^{(2)}(\tau)=1+\Delta c(\tau)\, ,
\end{equation}
where the correction $\Delta c(\tau)$ comes from the Wick decomposition of magnon correlators. As discussed above, we focus on equal-time ($\tau=0$) correlations in the steady-state, abbreviating $C \equiv C(0)$ and $\Delta c(0)\equiv\Delta c$.

The current-current cross correlation function is calculated perturbatively in $J$~\footnote{In nanojunctions in the \textit{ballistic} transport regime, electron cross-correlations are directly related to the junction conductance by fluctuation-dissipation because of charge conservation \cite{Cottet:2004io,*Souza:2008hb,*Blanter2000}. Such a relation applies to the magnonic spin current noise across one interface~\cite{Kamra2016,*Kamra2017}, but not to the cross-correlated spin current. Across the FI, however, we expect dephasing by magnon-phonon interactions to invalidate this relation, justifying a perturbative treatment where only the first few orders of $J$ are kept.}. Such an approach is consistent with the idea that the metallic leads act as weak probes of the magnet, thereby preserving a well-defined notion of a magnon rather than an entangled magnon/electron-hole excitation. Relegating the detailed calculation to the supplemental material, we summarize our results here.

Because the lowest nonvanishing contribution to $\mathcal{C}$ is fourth order in $J$, the lowest order terms (which we consider here) each contain four magnon operators, or two factors of the magnon density. There are therefore three types of terms: those $\sim N_c^2$, those $\sim N_i^2$, and those $\sim N_i N_c$, which we denote as $C_{cc}$, $C_{ii}$, and $C_{ic}$, respectively. One thus has:
\begin{equation}
C=C_{cc}+C_{ii}+C_{ic}
\end{equation}
where $C_{ii}=C_{ii}^{(\rm{X})}+C_{ii}^{(\parallel)}$ and $C_{ic}=C_{ic}^{(\rm{X})}+C_{ic}^{(\parallel)}$, with the superscripts $(\rm{X})$ and $(\parallel)$ respectively denoting diagrams that cross and do not cross the FI (see Supplemental Material). As shown in the Supplemental Material, one finds: $C_{cc}=I^{(c)}_L I^{(c)}_R$, $C_{ii}^{(\parallel)}=I^{(i)}_L I^{(i)}_R$, and $C_{ic}^{(\parallel)}=I^{(c)}_L I^{(i)}_R+I^{(i)}_L I^{(c)}_R$, where $I^{(i)}_l=2\pi D^2 J^2 E_m (N_i-N_{\mathrm{NM}})$ and $I^{(c)}_l=2\pi D^2 J^2 E_m N_c$ are the incoherent and coherent contributions to the $l=L,R$ spin current $I_l=I^{(i)}_l+I^{(c)}_l$. The sum of $C_{cc}$ and the uncrossed terms is thus equal to the product of the currents: $C_{cc}+C_{ii}^{(\parallel)}+C_{ic}^{(\parallel)}=I_L I_R$. One thus obtains $\Delta c=(C^{(\rm{X})}_{ii}+C^{(\rm{X})}_{ic})/I_L I_R$, so that $c^{(2)}$ is obtained directly by evaluating the crossed diagrams.

%When magnons are purely coherent, the crossed diagrams vanish, $\Delta c=0$, and $c=1$.

Next, let us consider the cross correlations under various scenarios.  First, in equilibrium, $N=N_{\mathrm{NM}}$, and the currents $I_L$ and $I_R$ vanish, as does coherent dynamics ($\varphi=0$), so $C_{cc}=C_{ic}=0$. At finite temperature, however, incoherent magnons are excited by thermal fluctuations. While the uncrossed diagrams vanish ($C^{\mathrm{(\parallel)}}_{ii}=I_L I_R=0$), the crossed diagrams $C^{\mathrm{(\rm{X})}}_{ii}$ do not, reflecting correlations between fluctuations of the left and right currents. Thus, in equilibrium, the normalized correlation coefficient $c^{(2)}=1+\Delta c$ diverges.

Second, consider heating of the FI. Under heating, a spin current flows from the FI to the normal metal leads.  All of the coherent terms, which are $\propto N_c$, are zero, leaving $C=C_{ii}=C_{ii}^{(\rm{X})}+C_{ii}^{(\parallel)}$.  As can be seen in the left-hand side Fig.~\ref{results}, as $T_m$ increases from $T$, $I_L=I_R$ becomes nonzero, decreasing $c^{(2)}$ from infinity. As shown in the Supplemental Material, when $T_m$ is sufficiently large that $N_m\gg N$, one finds after some work that $C_{ii}^{(\mathrm{X})}\rightarrow I_L I_R$, and therefore $c^{(2)}\rightarrow 2$. Under strong incoherent driving, thermal magnons dominate the spin currents and their correlations, and $c^{(2)}\sim \langle \hat{\varphi}^\dagger \hat{\varphi}^\dagger \hat{\varphi}  \hat{\varphi}\rangle/|\hat{\varphi}^\dagger \hat{\varphi}|^2=2$, reflecting bunching of thermal magnons~\footnote{Note that under cooling of the FI, however, fluctuations of electron-hole excitations dominate over those of magnons; as the cross-correlations do not capture thermal bunching of electron-hole pairs, we do not find that $c^{(2)}$ saturates at $2$ under cooling.}. Note also that because of \textit{quantum} fluctuations, even as the metal leads temperature $T$ is reduced to zero, a sufficiently large bias $\Delta T=T_m-T$ is required in order to observe thermal magnon bunching, i.e. $c^{(2)}=2$.

Last, consider driving by FMR.  Here, only coherent magnons are excited by the external field, and the spin currents $I_L=I^{(c)}_L=I_R=I^{(c)}_R$ are determined by the FMR power. Because the incoherent spin currents are zero ($N_i=N_{\mathrm{NM}}$), $C_{ii}^{(\parallel)}=C_{ic}^{(\parallel)}=0$, but the crossed terms $C^{(\rm{X})}_{ii}$ and $C^{(\rm{X})}_{ic}$ survive, reflecting correlations that arise from spin cotunneling.  At sufficiently large FMR power, however, $N_c \gg N_i$, and $C_{cc}=I_L I_R \gg  C^{(\rm{X})}_{ii},C^{(\rm{X})}_{ic}$, and therefore $\Delta c\rightarrow 0$, so $c^{(2)} \rightarrow 1$. Thus, at sufficiently large FMR power, the spin current correlations are dominated by the coherent magnons, and $c^{(2)}\approx g^{(2)}= \langle \varphi^* \varphi^*\varphi  \varphi\rangle/|\varphi^*\varphi|^2=1$. 
Note that even as $T\rightarrow 0$, $c^{(2)}$ is not necessarily equal to 1 at finite $N_c$. As with temperature biasing, \textit{quantum} fluctuations of magnons and electron/hole pairs require that additionally $N_c \gg 1$ in order for $c^{(2)}$ to saturate at 1.

{\it Measurement of cross correlations.} --- In the present device geometry, the spin current cross correlation function $C$ can be obtained from the spin current fluctuations $\mathcal{S}(\tau)\equiv(1/2)\langle \{\delta \hat{I}_L(t),\delta \hat{I}_R(t+\tau)\}\rangle$, where $\delta \hat{I}_l\equiv\hat{I}_l-\langle \hat{I}_l \rangle$\ak{.} In turn, $\mathcal{S}(\tau)$ is obtained by fourier transforming the power spectral density $\mathcal{S}(\omega)$. It is straightforward to see that $C=I_L I_R+\mathcal{S}(\tau=0)$, so $\Delta c=\mathcal{S}(\tau=0)/I_LI_R$. It should be noted that in an actual measurement, the detector bandwidth sets the lower limit on $\tau$ for which $\mathcal{S}(\tau)$ can be measured.  Thus, $\tau\rightarrow 0$ corresponds to $\tau$ approximately approaching the inverse of the detector bandwidth.

One might wonder why the spin current noise cannot alone be used to infer the coherence of the magnon state. The problem is that for both heating of the FI and FMR, $\mathcal{S}(0)$ increases with driving. In the former case, when the temperature difference between FI and leads is large enough, $\mathcal{S}(0)=C_{ii}^{(\mathrm{X})} \sim N_i^2$, while $I_L I_R\sim N_i^2$, so that their ratio $\Delta c$ saturates at constant value.  In the latter case, the noise $\mathcal{S}(0)$ grows with $C^{\mathrm{(X)}}_{ic} \sim N_c$, but because $I_L I_R\sim N_c^2$, $\Delta c\sim 1/N_c$ approaches zero with increasing FMR power.  Thus, in order to distinguish the cross-correlations of coherent magnons from those of incoherent magnons, it is expedient to use the normalized quantity $c^{(2)}$, rather than simply the noise $\mathcal{S}$. 

%\section{discussion}

%\ak{ In the present device geometry, an experimental method to measure $c^{(2)}(\tau)$ involves direct measurement of the power spectral density $\mathcal{S}(\omega)$ of the spin current cross-correlation~\cite{Kamra2014}: $\mathcal{S}(\omega) = \int d\tau C^{(2)}(\tau) \exp{(i \omega \tau)}$, whence we obtain:
%\begin{align}
%C^{(2)}(\tau) = & \frac{1}{2 \pi} \int d \omega \mathcal{S}(\omega) e^{ - i \omega \tau}.
%\end{align}
%Here, $C^{(2)}(\tau) \equiv \left \langle  \hat{I}_L(t) \hat{I}_R(t + \tau)  \right \rangle$. The normalized cross-correlation can be obtained from the knowledge of $C^{(2)}(\tau)$ and the dc spin currents $I_{L,R} \equiv \langle \hat{I}_{L,R} \rangle$. For time invariant systems, the physical quantities do not depend on $t$. When the time invariance is broken, for example by a time dependent applied magnetic field, the corresponding quantities are obtained by averaging over $t$~\cite{Kamra2016B,Blanter2000}. In most cases of interest, this more precise definition including averaging over $t$ leads to the same result as the definitions employed herein~\cite{Kamra2016B}. In an actual measurement, the detector bandwidth sets the lower limit on $\tau$ for which $C^{(2)}(\tau)$ can be measured. Thus, $\tau \to 0$ corresponds to $\tau$ approaching inverse of the detector bandwidth, approximately.

{\it Discussion.} --- We have focused on a simple model to capture the essential physics of magnon bunching; we now comment on some of the approximations, along with their validity, employed in our analysis. First, we have assumed that metal contacts are placed in close proximity to one another.  If the contacts are too far apart so that the thermal wavelength of magnons is much smaller than the propagation length, scattering of magnons may destroy correlations between interfaces, reducing $c^{(2)}$ from 2 in the case of strong driving of incoherent magnons.  If, on the other hand, the FI is too thin or the contacts too close, tunneling of electrons across the FI becomes possible, giving rise to a spin-polarized charge current that distorts the magnonic signal. Second, the actual observation of pure spin currents and their fluctuations require conversion to measurable charge currents by, for example, the inverse spin Hall effect; this process may introduce additional charge current noise that is convoluted with that of cross-correlations.  Third, we have considered a macrospin model, wherein higher energy magnons are gapped out.  While this may be a reasonable assumption for small structures at low temperatures, clearly as the FI is heated, micromagnetic modes must be taken into account. 

%---------------------------------------------Summary------------------------------------------- %

{\it Summary.} --- We have proposed the normalized spin-current cross-correlation as a measure of magnon coherence in magnet/metal hybrid systems. Evaluating it for zero time delay between the correlated spin currents, we have shown that it approaches 2 for a thermal magnon state thereby demonstrating magnon-bunching. It approaches 1 when the magnon mode is in a coherent state, thereby serving as means to characterize the that state. While we have presented a proof-of-principle analysis highlighting the potential of the normalized spin-current cross-correlation, its demonstrated similarity to the optical coherence function paves the way for its usefulness to the field of quantum magnonics.

%---------------------------------------------Acknowledgments------------------------------------------- %

{\it Acknowledgments.} We thank Asle Sudb{\o} and Arne Brataas for valuable discussions. S. B. and R. D. are supported by funding from the Stichting voor Fundamenteel Onderzoek der Materie (FOM) and the European Research Council via Consolidator Grant number 725509 SPINBEYOND.  W. B. and A. K. are funded by the DFG through SFB 767 and the Alexander von Humboldt Foundation. R. D. and A. K. also acknowledge financial support from the Research Council of Norway through its Centers of Excellence funding scheme, project 262633, ``QuSpin''.
%---------------------------------------------Bib------------------------------------------- %

%

%\bibliography{g2BIB}


%---------------------------------------------Supp Material------------------------------------------- %

\begin{widetext}
\section{Calculation of cross-correlations}
In this section, we give a detailed calculation of the current-current correlator
\begin{equation}
C(t,t')\equiv (1/2)\langle \hat{I}_L(t)  \hat{I}_R(t') + \hat{I}_R(t')  \hat{I}_L(t) \rangle \, .
\end{equation}
We perform a perturbative treatment in $\hat{H}_J$ to lowest nonvanishing (i.e. second) order, yielding a contribution to $C(t,t')$ of order $J^4$. It is, in this case, expedient to employ contour-ordered Green's functions.  The real-time correlation $C(t,t')$ may be obtained from  the contour-time object $\mathcal{C}(\tau,\tau')=(1/2) \langle \hat{I}_L(\tau ) \hat{I}_R(\tau')\rangle$ (with $\tau$ denoting contour-times, not to be confused with $t-t'$ used in the main paper):
\begin{equation}
C(t,t')=\mathcal{C}^K(t,t')\, ,
\end{equation}
where the superscript $K$ denotes the Keldysh or kinetic component: $\langle \hat{A}(t)  \hat{B}(t')\rangle^K=\langle \hat{A}(t)  \hat{B}(t')\rangle+\langle \hat{B}(t') \hat{A}(t)  \rangle$.

To lowest nonzero order (i.e. $J^4$) in perturbation theory with respect to $\hat{\mathcal{H}}_J$, one finds:
\begin{align}
\mathcal{C}(\tau,\tau')=\frac{1}{2}\langle  T_c \left[e^{-i\int_c d \tilde{\tau } \hat{\mathcal{H}}_J(\tilde{\tau})} \hat{I}_L(\tau)  \hat{I}_R(\tau') \right]\rangle_0 \nonumber\\
\approx-\frac{1}{4}\langle  T_c \left[\int_c d \tau_1   \int_c d \tau _2 \hat{\mathcal{H}}_J (\tau _1) \hat{\mathcal{H}}_J (\tau _2)  \hat{I}_L(\tau)  \hat{I}_R(\tau') \right]\rangle_0+&\mathcal{O}(J^6)
\label{Stt}
\end{align}
Here, $c$ is the time-contour and $T_c$ is the contour-ordering; the contour-time dependence of all of the operators in Eq.~(\ref{Stt}) are understood to be in interaction picture with respect to $\hat{\mathcal{H}}_0=\hat{\mathcal{H}}_m+\hat{\mathcal{H}}_L+\hat{\mathcal{H}}_R$.  Inserting $\hat{\mathcal{H}}_J$ in Eq.~(\ref{Stt}) and working out the products, one obtains: 
\begin{equation}
\mathcal{C}(\tau,\tau')=\mathrm{Re}[\mathcal{C}_1(\tau,\tau')-\mathcal{C}_2(\tau,\tau')]
\label{12def}
\end{equation}
where
\begin{align}
\mathcal{C}_1(\tau,\tau')=\langle  T_c [\int_c d \tau_1   \int_c d \tau _2 \hat{L}^\dagger (\tau _1) \hat{R}^\dagger (\tau _2)  \hat{L}(\tau)  \hat{R}(\tau') ]\rangle_0\ ,
\nonumber\\
\mathcal{C}_2(\tau,\tau')=\langle  T_c [\int_c d \tau_1   \int_c d \tau _2 \hat{L}^\dagger (\tau _1) \hat{R} (\tau _2)  \hat{L}(\tau)  \hat{R}^\dagger(\tau') ]\rangle_0
\nonumber\,.
\end{align}
Using $\hat{L}\equiv \sum_{ K}J_{k} \hat{L}_K\hat{\phi}^\dagger$, these can be more explicitly written:
\begin{align}
\mathcal{C}_1(\tau,\tau')=\nonumber
\sum_{K_1K_2KK'} J_{K_1}^* J_{K_2}^*J_{K}J_{K'}
\int_c d\tau_1 \int_c d\tau_2 G_L(1,1_1) G_R(1',1_2) G_m(1_1,1_2;1,1')
\end{align}
\begin{align}
\mathcal{C}_2(\tau,\tau')=\nonumber
\sum_{K_1K_2KK'} J^*_{K_1} J_{K_2}J_{K}J^*_{K'} \int_c d\tau_1 \int_c d\tau_2 G_L(1,1_1) G_R(1_2,1') G_m(1_1,1';1,1_2)\, ,
\end{align}
where
\begin{equation}
G_L(1,1')\equiv-i\langle T_c[\hat{L}_K(\tau)\hat{L}^\dagger_{K'}(\tau')]\rangle_0
\end{equation}
is the contour-ordered Green's function for LNM electron-hole pairs, with $1=(K,t)$ as a collective index;  a similar definition for the RNM electron-hole function $G_R(1,1')$ holds. The quantity:
\begin{equation}
G_m(1_a,1_b;1_c,1_d)=\langle T_c[\hat{\phi}(\tau_a)\hat{\phi}(\tau_b)\hat{\phi}^\dagger(\tau_c) \hat{\phi}^\dagger(\tau_d)]\rangle_0
\label{fourpoint}
\end{equation}
depends crucially on whether the magnon operators are coherent or incoherent. For this reason, we obtain three types of contributions to $\mathcal{C}_1$ and $\mathcal{C}_2$: terms that involve only incoherent operators ($\mathcal{C}_{n,ii}$ for $n=1,2$), terms that involve only coherent magnon operators ($\mathcal{C}_{n,cc}$) and cross-terms that involve both ($\mathcal{C}_{n,ic}$):
\begin{equation}
\mathcal{C}_n= \mathcal{C}_{n,ii}+\mathcal{C}_{n,ic}+\mathcal{C}_{n,cc}\, .
\end{equation}

\subsection{Purely Incoherent Terms: $ \mathcal{C}_{n,ii}$}
\label{iisec}
In this subsection, we calculate the purely incoherent terms, $\mathcal{C}_{n,ii}$. Because for these terms, $\hat{\phi}=\hat{\varphi}$, which is assumed to be given by a Gaussian average at zero$^{th}$ order in coupling, we may decompose the four-point contour-ordered magnon correlator into two terms via Wicks theorem:
\begin{align}
\nonumber
 G_m(1_a,1_b;1_c,1_d)=-G_m(1_a,1_c)&G_m(1_b,1_d)-G_m(1_a,1_d)G_m(1_b,1_c)\, ,
 \end{align}
 where $G_m(1_a,1_c)=-i\langle T_c[\hat{\phi}(\tau_a)\hat{\phi}^\dagger(\tau_c)] \rangle_0$ is the contour-ordered magnon Green's function.

 This gives rise to two terms in each $\mathcal{C}_{n(ii)}$:
\begin{equation}
\mathcal{C}_{n,ii}=\mathcal{C}^{(\parallel)}_{n,ii}+\mathcal{C}^{(\rm{X})}_{n,ii}
\label{incoh}
\end{equation}
for $n=1,2$ where
\begin{align}
\mathcal{C}^{(\parallel)}_{1,ii}(\tau,\tau')=
-\sum_{K_1K} J_{K_1}^* J_{K}\int_c d\tau_1  G_L(1,1_1) G_m(1_1,1)
 \sum_{K_2K'}J_{K_2}^*J_{K'}\int_c d\tau_2 G_R(1',1_2) G_m(1_2;1')
\label{s1p}
\end{align}
\begin{align}
\mathcal{C}^{(\parallel)}_{2,ii}(\tau,\tau')=
-\sum_{K_1K} J_{K_1}^* J_{K}\int_c d\tau_1  G_L(1,1_1) G_m(1_1,1)  \sum_{K_2K'}J_{K_2}J^*_{K'}\int_c d\tau_2 G_m(1',1_2) G_R(1_2;1')
\label{s2p}
\end{align}
are the ``uncrossed'' terms.  The ``crossed" terms are:
\begin{align}
\mathcal{C}^{(\rm{X})}_{1,ii}(\tau,\tau')=
-\sum_{K_1K_2KK'} J_{K_1}^* J_{K_2}^*J_{K}J_{K'}
\int_c d\tau_1 \int_c d\tau_2 G_L(1,1_1) G_m(1_1,1')G_R(1',1_2) G_m(1_2;1)
\label{s1x}
\end{align}
and
\begin{align}
\mathcal{C}^{(\rm{X})}_{2,ii}(\tau,\tau')=
-\sum_{K_1K_2KK'} J^*_{K_1} J_{K_2}J_{K}J^*_{K'}
\int_c d\tau_1 \int_c d\tau_2 G_L(1,1_1) G_m(1_1,1_2)G_R(1_2,1') G_m(1';1)
\label{s2x}
\end{align}
The diagrams for these are given depicted in Fig.~\ref{iidiagrams}.

\begin{figure}[pt]
\includegraphics[width=0.6\linewidth,clip=]{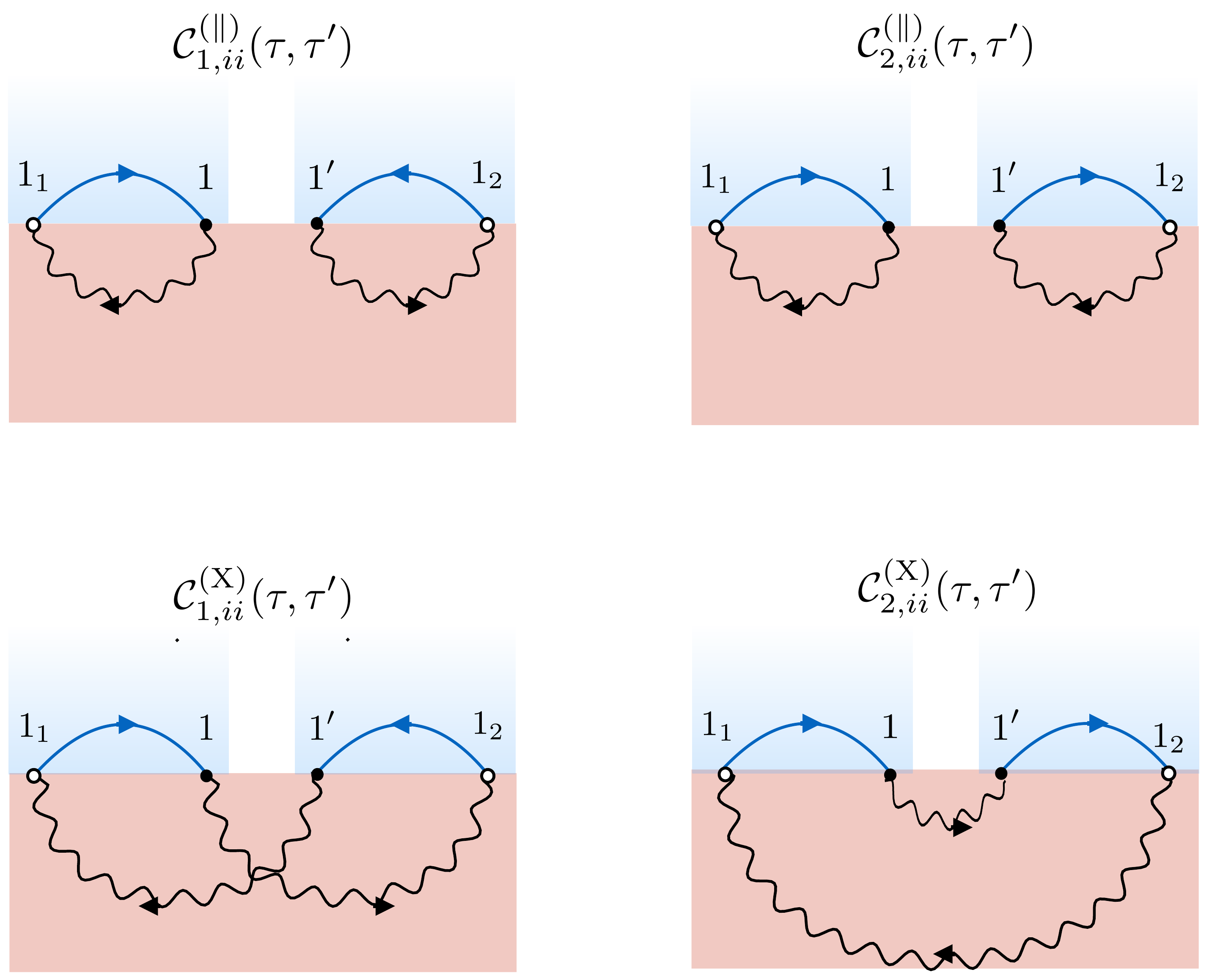}
\caption{Contour-ordered Feynman diagrams for uncrossed (top two) and crossed (bottom two) contour-ordered noise correlators, corresponding to Eqs.~(\ref{s1p})-(\ref{s2x}). Wavy lines represent magnon Green's functions, $G_m(1,2)$, while solid lines electron-hole Green's functions, $G_l(1,2)$ for $l=L,R$. Solid black points are external vertices, while white points with a black outline are internal indices.  The real time Keldysh components are given by Eqs.~(\ref{s1xii}) and~(\ref{s2xii}).}
\label{iidiagrams}
\end{figure} 

Next we take the Keldysh components of Eqs.~(\ref{s1p})-(\ref{s2x}). We break the resulting expressions expressions into real-time greater than, less than, retarded and advanced Green's functions of noninteracting electrons and magnons via the Langreth relations. After some work, the uncrossed contributions are found to be (see Eq.~(\ref{12def}))
\begin{equation}
C^{(\rm{\parallel})}_{ii}(t,t)=[\mathcal{C}^{(\rm{\parallel})}_{ii}(t,t)]^K=\lim_{t\rightarrow t'}\mathrm{Re}[\mathcal{C}^{(\rm{\parallel})}_{1,ii}(t,t')-\mathcal{C}^{(\rm{\parallel})}_{2,ii}(t,t')]^K=I_L^{(i)}I_R^{(i)}
\label{inunc}
\end{equation}
where $I_l^{(i)}=2\pi D^2  J^2 E [N^{(i)}-n^{(l)}_{B}\left(\epsilon_q \right)]\, $
is the incoherent contribution to the spin current through the $l=L,R$ interfaces. For generality, we include nonequilibirium spin-dependent splitting $\mu_{l\sigma}$ in the leads (which are set to zero in the main text), so $\langle \hat{b}_{lk\sigma}^\dagger \hat{b}_{l'k'\sigma'}\rangle_0=\delta_{ll'}\delta_{kk'}\delta_{\sigma\sigma'} n^{(l)}_{F ,\sigma}(\epsilon_{k})\,$, with $n^{(l)}_{F,\sigma }(\epsilon)=[1+e^{(\epsilon-\mu_{l\sigma})/T_l}]^{-1}$, and $T_l$ the temperature in lead-$l$.

The crossed, incoherent contributions are:
\begin{equation}
C^{(\rm{X})}_{ii}(t,t)=\lim_{t\rightarrow t'}\mathrm{Re}[\mathcal{C}^{(\rm{X})}_{1,ii}(t,t')-\mathcal{C}^{(\rm{X})}_{2,ii}(t,t')]^K
\label{incohcrossed}
\end{equation}

where, using $J^2\equiv \sum_{K} \left| J_{K}\right|^2 \delta(\epsilon_k-\epsilon_F) \delta(\epsilon_{\tilde{k}}-\epsilon_F)/D^2$:
\begin{align}
 \lim_{t\rightarrow t'}[\mathcal{C}^{(\rm{X})}_{1,ii}(t,t')]^K=-2 D^4J^4 \int d\xi(\xi-\mu_{L})\frac{N_i-n_{B}^{(L)}\left(\xi\right)}{\left(E-\xi+i0^+\right)}\int d\xi'(\xi'-\mu_{R})\frac{N_i-n_{B}^{(R)}\left(\xi'\right)}{\left(E_{q}-\xi'+i0^+\right)}
 \label{s1xii}
 \\
  \lim_{t\rightarrow t'}[\mathcal{C}^{(\rm{X})}_{2,ii}(t,t')]^K=D^4 J^4 \int d\xi\int d\xi'\frac{(\xi-\mu_{L})(\xi'-\mu_{R})}{\left(E-\xi'-i0^+\right) \left(\xi'-\xi+i0^+\right)} \left[n^{(L)}_{B}(\xi)\left(1+N_i\right)+\left(1+n^{(L)}_{B}(\xi)\right)N_i \right]
  \nonumber \\
  +D^4J^4\int d\xi\int d\xi'\frac{(\xi-\mu_{L})(\xi'-\mu_{R})}{\left(E-\xi+i0^+\right) \left(\xi-\xi'-i0^+\right)} \left[n^{(R)}_{B}(\xi')\left(1+N_i\right)+\left(1+n^{(R)}_{B}(\xi')\right)N_i \right] \nonumber\\
  -D^4J^4\int d\xi\int d\xi'\frac{(\xi-\mu_{L})(\xi'-\mu_{R})}{\left(\xi-E-i0^+\right) \left(\xi'-E+i0^+\right)} \left[N_i\left(1+N_i\right)+\left(1+N_i\right)N_i \, .\right]&
  \label{s2xii}
  \end{align}
   Here, $\xi=\xi_\uparrow-\xi_\downarrow$ and $\xi'=\xi'_\uparrow-\xi'_\downarrow$ are electron-hole energies (i.e. the energy differences between spin-up and spin-down electrons), with $n^{(l)}_B(\epsilon)\equiv[e^{(\epsilon-\mu_l)/T_l}-1]$ and $\mu_l \equiv \mu_{l \uparrow} -\mu_{l \downarrow}$.
% Finally, if we employ the Plemelj relation, $1/(\omega \pm i 0^+)=\mp i\pi \delta(\omega)+\mathcal{P}(1/\omega)$, where $\mathcal{P}$ denotes principle value, and assuming that you can drop the contributions from the principle value terms, one obtains:
% \begin{widetext}
 %\begin{equation}
%\boxed{ \lim_{t\rightarrow t'}\Re[\mathcal{S}^{(ii)}_{1\rm{X}}(\tau,\tau')]^K=\frac{1}{2}I^{(i)}_L I^{(i)}_R} 
%\label{1Xafterplemelj}
%\end{equation}
%\begin{equation}
%\boxed{\lim_{t\rightarrow t'}\Re[\mathcal{S}^{(ii)}_{2\rm{X}}(\tau,\tau')]^K=\sum_{qq'}D_L^2 D_R^2 \pi^2 (E_{q'}-
%\mu_L)(E_{q'}-\mu_R) \left[\left(N_{q'}^{\left(x\right)}-N_{q'}^{\left(i\right)}\right)\left(1+N_{q}^{\left(i\right)}\right)+\left(1+N_{q'}^{\left(x\right)}-N_{q'}^{\left(i\right)}\right)N_{q}^{(i)}\right]}
%\label{2Xafterplemelj}
% \end{equation}
 % \end{widetext}
 % where $N_{q'}^{\left(x\right)}\equiv \sum_{l=L,R}n^{(l)}_{B}(E_{q'})$ correspond to electron-hole excitations of the left and right leads, with, again, $n^{(l)}_{B}(\epsilon)=1/(e^{(\epsilon-\mu_L )/T_l}-1) $. 
 
 Together, Eqs.~(\ref{totcurrent}),~(\ref{incoh}),~(\ref{inunc}),~(\ref{incohcrossed}),~(\ref{s1xii}) and~(\ref{s2xii}) allow us to compute the correlator $\mathcal{C}_{ii}$, for an arbitrary number $N_i$ of purely incoherent magnons.  Using Eqs~(\ref{s1xii}) and~(\ref{s2xii}), it is straightforward to show that under strong heating of the FI, where $N_i \gg n_B^{(l)}(\xi)=N_{\mathrm{NM}}$, 
 \begin{equation}
C^{(\rm{X})}_{ii}(t,t) \rightarrow \frac{}{}I_L^{(i)}I_R^{(i)}\, ,
 \label{ciiX}
 \end{equation}
 thus establishing $c^{(2)}\rightarrow 2$ under heating of the FI.

  \subsection{Purely Coherent Terms: $ \mathcal{C}_{n,cc}$}
  \label{seccc}
  These are terms in which all four of the magnon operator in Eq.~(\ref{fourpoint}) is a scalar: $\hat{\phi}\rightarrow  \varphi$.  In this case, the four-point magnon function in Eq.~(\ref{fourpoint}) does not factorize into two pieces, and the diagramatics simplify considerably (see Fig.~\ref{coherentdiagrams}).  Following a similar procedure as above, one finds:
  \begin{equation}
\mathcal{C}_{cc}(t,t)=\lim_{t\rightarrow t'}\Re\left[\mathcal{C}_{1,cc}(t,t')-\mathcal{C}_{2,cc}(t,t') \right]^K=I_L^{(c)} I_R^{(c)}\, ,
\label{cc}
\end{equation}
where
\begin{equation}
I_l^{(c)}=2\pi D^2  J ^2 \left(E -\mu_l\right) N_c\, ,
\label{ccurrent}
\end{equation}
is the contribution of the coherent current to the total spin current through the $l=L,R$ interface. 

\begin{figure}[pt]
\includegraphics[width=0.6\linewidth,clip=]{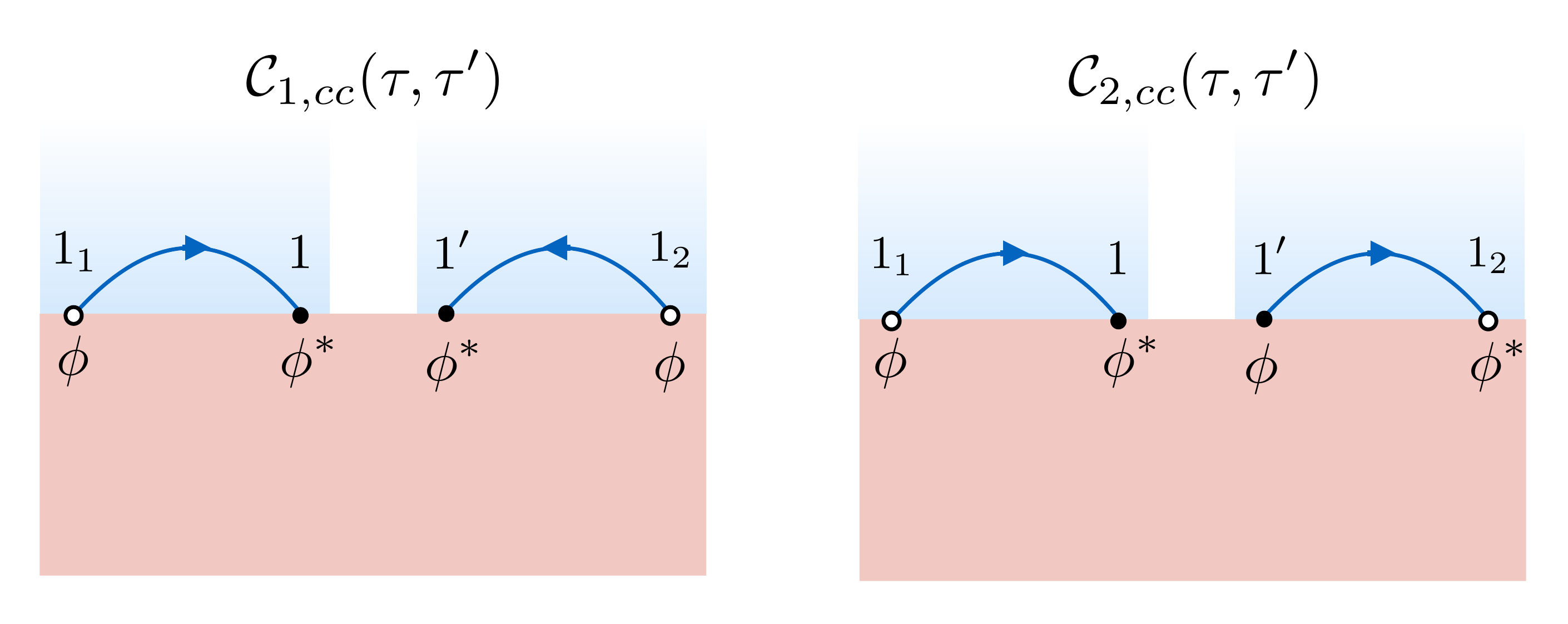}
\caption{Contour-ordered Feynman diagrams for purely coherent terms. }.
\label{coherentdiagrams}
\end{figure} 

  \subsection{Mixed Terms: $\mathcal{C}_{n,ic}$}
 \label{secic}

  These are terms in which two of the magnon operators in Eq.~(\ref{fourpoint}) are c-numbers ($ \varphi$), while the other pair are operators ($\hat{\varphi}$). These can are obtained from a similar procedure as above, resulting in:
    \begin{equation}
 \mathcal{C}_{(ic)}(t,t)=\mathcal{C}^{(\parallel)}_{(ic)}(t,t)+\mathcal{C}^{(\rm{X})}_{(ic)}(t,t)\label{ic1}
\end{equation}
 with uncrossed terms: \begin{equation}
\mathcal{C}^{(\parallel)}_{(ic)}(t,t)=\lim_{t\rightarrow t'}\Re\left[\mathcal{C}^{(\parallel)}_{(1,ic)}(\tau,\tau')-\mathcal{C}^{(2,\parallel)}_{(ic)}(\tau,\tau') \right]^K=I_L^{(i)} I_R^{(c)}+I_L^{(c)} I_R^{(i)}
 \label{ic2}
\end{equation}
(see Fig.~\ref{icuncross}) and crossed terms:
\begin{equation}
\mathcal{C}^{(\rm{X})}_{(ic)}(t,t)=\lim_{t\rightarrow t'}\Re\left[\mathcal{C}^{(\rm{X})}_{(1,ic)}(t,t')-\mathcal{C}^{(\rm{X})}_{(2,ic)}(t,t') \right]^K
 \label{ipp2}
\end{equation}
 where
 \begin{align}
 \lim_{t\rightarrow t'}[\mathcal{C}^{(\rm{X})}_{(1,ic)}(t,t')]^K=-2D^4J^4 \int d\xi(\xi-\mu_{L}) \int d\xi'(\xi'-\mu_{R})\frac{N_{c}\left( N_{i}-n_{B}^{(R)}\left(\xi'\right)\right)+N_{c}\left( N_{i}-n_{B}^{(L)}\left(\xi\right)\right)}{\left(E_m-\xi+i0^+\right)\left(E_m-\xi'+i0^+\right)}
 \label{s1xic}
 \\
  \lim_{t\rightarrow t'}[\mathcal{C}^{(\rm{X})}_{(2,ic)}(t,t')]^K= D^4 J^4\int d\xi\int d\xi'\frac{(\xi-\mu_{L})(\xi'-\mu_{R})}{\left(E_m-\xi'-i0^+\right) \left(\xi'-\xi+i0^+\right)} \left(1+2 n^{(L)}_{B}(\xi)\right)N_c 
  \nonumber \\
  +D^4 J^4\int d\xi\int d\xi'\frac{(\xi-\mu_{L})(\xi'-\mu_{R})}{\left(E_m-\xi+i0^+\right) \left(\xi-\xi'-i0^+\right)} \left(1+2 n^{(R)}_{B}(\xi')\right)N_c  \nonumber\\
  -D^4 J^4\int d\xi\int d\xi'\frac{(\xi-\mu_{L})(\xi'-\mu_{R})}{\left(\xi-E_m-i0^+\right) \left(\xi'-E_m+i0^+\right)} \left[N_c\left(1+N_i\right)+N_i N_c+\left(1+N_i\right)N_c +N_cN_i\right]&\, ,
  \label{s2xic2}
  \end{align}
(see Fig.~(\ref{iccross})). This can also be obtained by sending $N_i \rightarrow N_c $ in the corresponding terms for $\mathcal{C}_{ii}$ above, and assuming that $N_c\gg 1$ so that $N_c+1\approx N_c$, i.e. that the coherent terms are not subject to quantum fluctuations. 

The curves in Fig. \ref{results} are obtained from numerical integration of Eqs.~(\ref{s1xii}),~(\ref{s2xii}),~(\ref{s1xic}) and~(\ref{s2xic2}) above, with energy cutoffs $\xi,\xi'=10 E_m$ in order to perform the integrations.
  
  \begin{figure}[pt]
\includegraphics[width=0.6\linewidth,clip=]{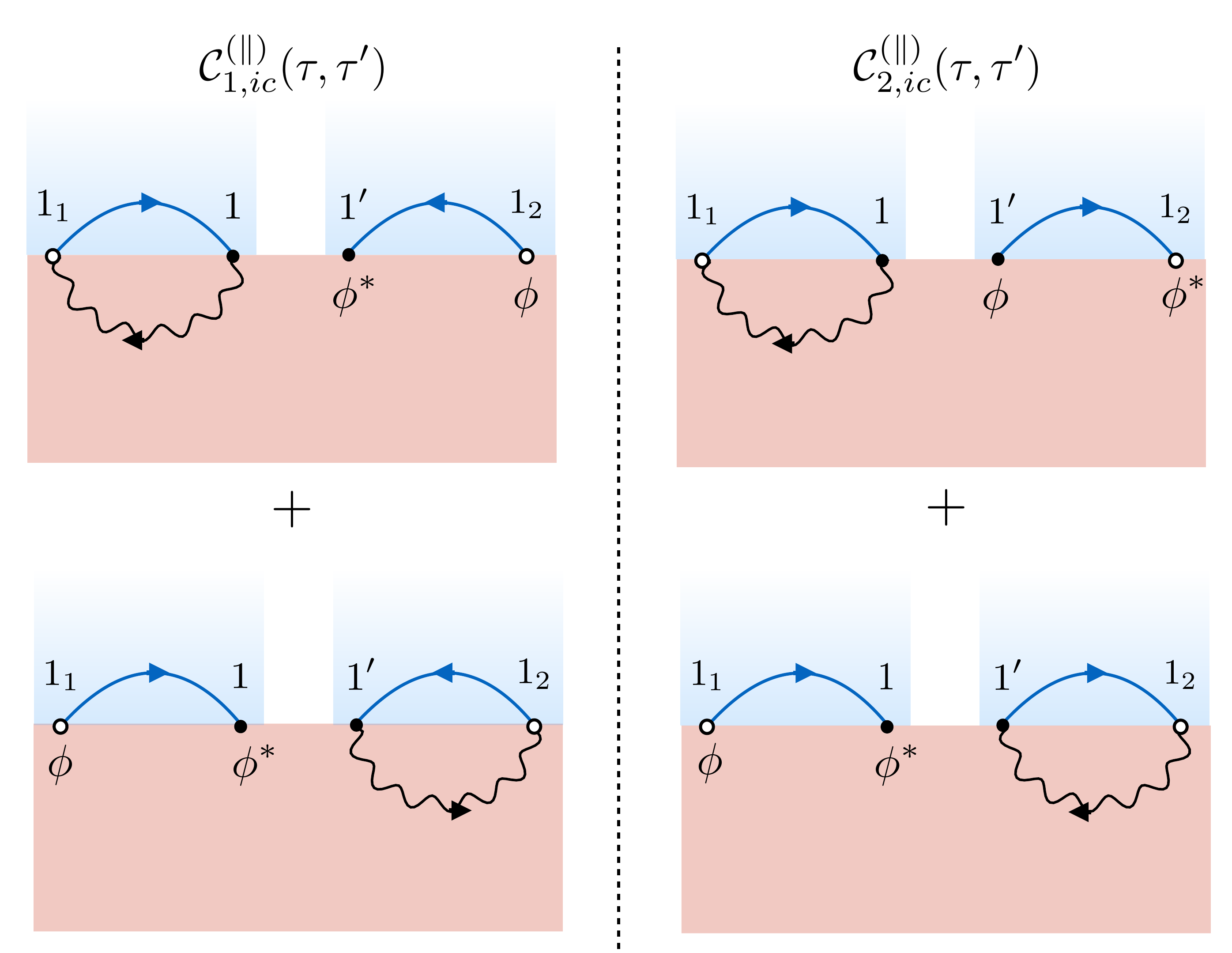}
\caption{Contour-ordered Feynman diagrams for mixed, uncrossed terms. }.
\label{icuncross}
\end{figure} 

\begin{figure}[pt]
\includegraphics[width=0.6\linewidth,clip=]{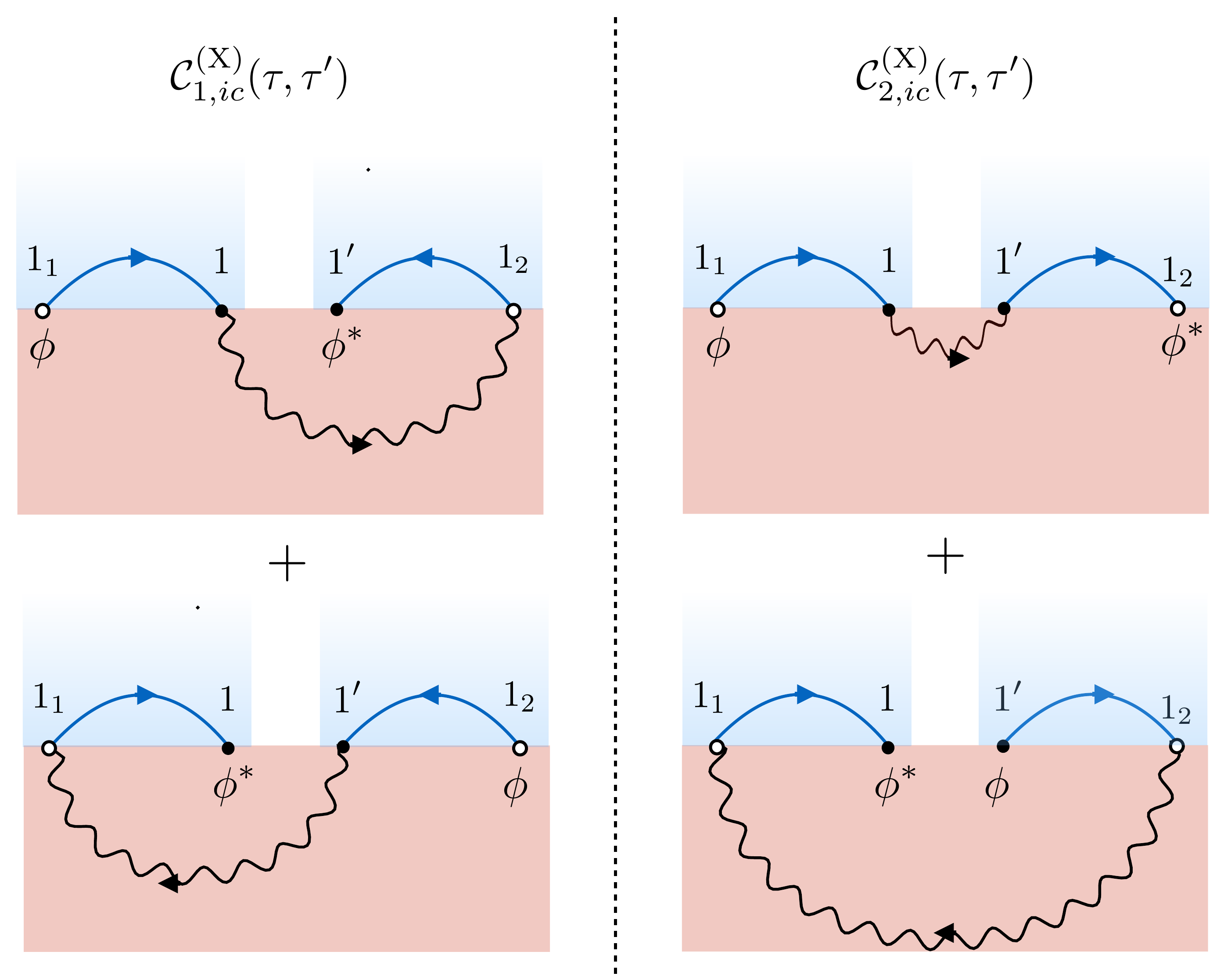}
\caption{Contour-ordered Feynman diagrams for mixed, crossed terms. }.
\label{iccross}
\end{figure}

\end{widetext}

\end{document}